\documentclass[10pt,aip, twocolumn,amssymb]{revtex4-1}   

\usepackage{amsmath}    
\usepackage{amsfonts}  
\usepackage{amssymb}
\usepackage{graphicx}   
\usepackage{multirow}
\usepackage{array}
\usepackage{tabulary}
\usepackage[utf8]{inputenc} 

\begin{document}

\title{ Controlling electronic structure through epitaxial strain in ZnSe/ZnTe nano-heterostructures}
\author{S. K. Yadav,$^{a,b}$} 
\email{syadav@lanl.gov, yadav.satyesh@gmail.com}   
\author{V. Sharma,$^{a}$ and R. Ramprasad$^{a}$}

\affiliation{$^{a}$ Materials Science and Engineering, University of Connecticut, Storrs, CT 06269, USA\\
$^{b}$ MST-8, Los Alamos National Laboratory, Los Alamos, NM 87545, USA}

\begin{abstract}

 Using first-principles computations, we study the effect of epitaxial strains on electronic structure variations across ZnSe/ZnTe nano-heterostructures. Epitaxial strains of various types are modeled using pseudomorphic ZnSe/ZnTe heterostructures. We find that a wide range of band gaps (spanning the visible solar spectrum) and band offsets (0-1.5 eV) is accessible across the heterostructures in a controllable manner via reasonable levels of epitaxial strain. In addition to quantum confinement effects, strain in ZnSe/ZnTe heterostructures may thus be viewed as a powerful degree of freedom that can enable the rational design of optoelectronic devices. 
\end{abstract}

\maketitle

\section{Introduction}

The control of electronic states in semiconductors through dilation of the crystal lattice was first explored in detail by Bardeen and Shockley.\cite{Bardeen-Sh} As the lattice contracts under hydrostatic strain, the splitting between the bonding valence states and anti-bonding conduction states increases, resulting in a net bandgap increase, and vice-versa. Similarly, under uniaxial and biaxial strains, first-principles calculations have predicted spectacular variation of the bandgap in bulk Zn\textit{X} (\textit{X} = O, S, Se, and Te)\cite{4,10} CdSe, CdTe,\cite{5} and GaN.\cite{6,7} Moreover, epitaxial strains in heterostructures strongly influence the variation of electronic properties across interfaces (e.g., the band offsets), reported in recent experimental works on ZnTe/ZnSe and CdTe/ZnSe core/shell quantum dots.\cite{b-1,b-2,b-3}

\begin{figure}
\includegraphics[,width=3.5in]{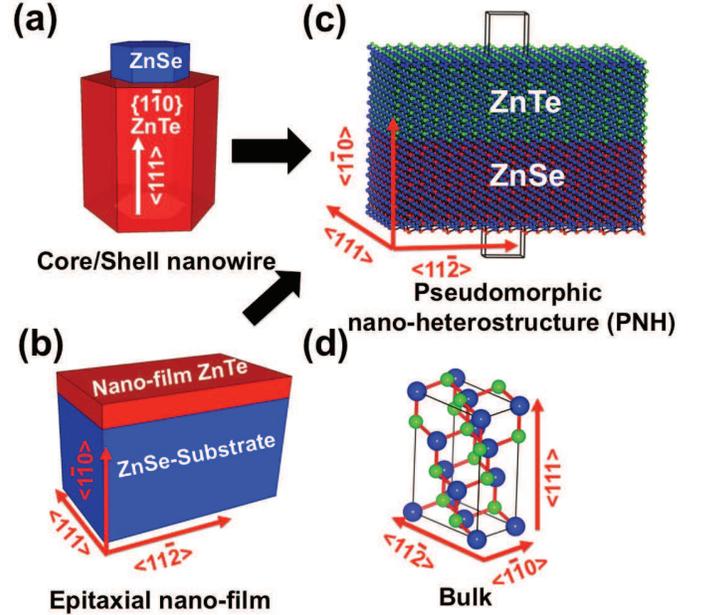}
\caption {(Color online) In-plane epitaxial strain in the $\{1\overline{1}0\}$ interfacial plane in (a) core/shell nanowire and (b) thin film grown on a substrate can be modeled using (c) pseudomorphic nano-heterostructure (PNH). (d) Bulk supercell is used for modeling strain in the bulk.}
{\label{sche}}
\end{figure}

\begin{figure}
\includegraphics[,width=2.5in]{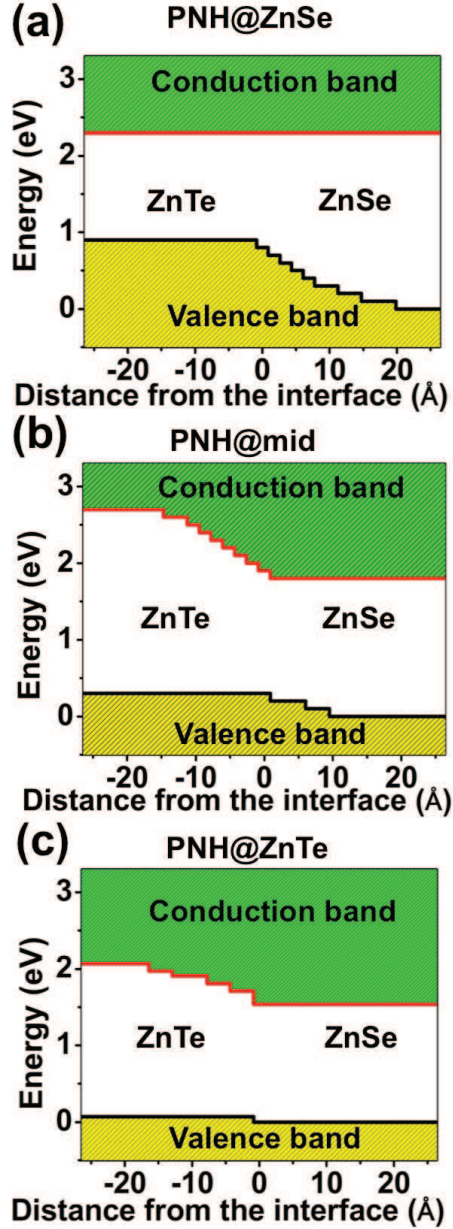}
\caption {(Color online) Conduction and valence band edges across the interface of PNHs calculated using layer decomposed density of states. Three configurations were considered: PNH@ZnSe, (b) PNH@mid and (c) PNH@ZnTe. The zero distance represents the interface, with ZnTe on left and ZnSe on right. The zero of energy is set to the lowest energy of the valence band edge. Calculated lattice constant for ZnSe and ZnTe are, 5.74 \AA and 6.18 \AA, respectively.}
{\label{energy}}
\end{figure}

In this work we explore the effect of epitaxial strains on electronic properties of ZnSe/ZnTe nano-heterostructures using first-principles calculations. The choice of ZnSe and ZnTe is motivated by the expected large strain-induced tunability of bandgaps of 2.71 eV and 2.39 eV, respectively, such as to span the visible solar spectrum.\cite{4} It is also likely that the 7.3\% lattice mismatch between these two systems may be accommodated in a coherent heterostructure, (as encountered in core/shell nanowire and epitaxial nano-films) due to increased yield strengths at nanometer length scales.\cite{opto-1,opto-2,opto-3,opto-4} Indeed, ZnSe/ZnTe heterostructures have recently been synthesized in the form of core/shell nanowires \cite{2-3} and quantum dots.\cite{b-2,b-3}

\section{Methods}

The effect of epitaxial strains in core/shell nanowire and nano-film on electronic properties of ZnSe/ZnTe heterostructures were modeled using pseudomorphic nano-heterostructures (PNHs), as shown in Fig.~\ref{sche}. Each layer parallel to the interface in PNHs have the same lattice spacing. Fig.~\ref{sche}(c) shows the atomic structure of a PNH used for modeling epitaxial strain in the $\{1\overline{1}0\}$ plane. The structure in the figure is obtained by periodically repeating a slab supercell with 16 layers each of ZnSe and ZnTe with vacuum of 10 \AA. 16 layers were sufficient to separate out the effect of quantum confinement, more layers would not affect the results. In experiments more layers would mean loss of coherency and formation of misfit dislocations at the interfaces. The model represents epitaxially strained situations in coherent core/shell nanowires grown along the $\textless111\textgreater$ direction with a $\{1\overline{1}0\}$ plane termination and coherent nano-film on a substrate grown along the $\{1\overline{1}0\}$ direction. Similarly, PNHs with $\{111\}$ and $\{11\overline{2}\}$ interfacial planes were used to model epitaxial strains in the respective planes.

To separate out the effects of quantum confinement and interfaces from the effect of strains on the electronic properties of ZnSe and ZnTe in PNHs, we also modeled bulk ZnSe and ZnTe under similar states of strains encountered in PNHs. Calculations involving strains required a supercell containing six pairs of ZnSe or ZnTe unit in zinc blende crystal structure Fig.~\ref{sche}(d).\cite{s-t} For example, to model the state of strain in ZnSe or ZnTe in the PNH in Fig.~\ref{sche}(c), the in-plane lattice parameter in the $\{1\overline{1}0\}$ plane is fixed to a desired value and the lattice parameter in the $\textless1\overline{1}0\textgreater$ direction is allowed to change so that there in no stress in the $\textless1\overline{1}0\textgreater$ direction.

Our first-principles density functional theory (DFT) calculations were performed using the Vienna \textit{ab initio} simulation package (VASP).\cite{14} Geometry optimizations were performed using the Perdew, Burke, and Ernzerhof (PBE) semilocal exchange-correlation functional\cite{12} and the projector-augmented wave methodology.\cite{13} The optimized geometries were subsequently used to determine the electronic structure using hybrid DFT calculations utilizing the specific functional referred to as the HSE06 functional in the literature.\cite{16} This functional is created by starting with the PBE exchange-correlation functional and replacing 25\% of the PBE exchange interaction by a screened nonlocal functional with an inverse screening length of 0.2 \AA$^{-1}$. A 6x4x3 Monkhorst-Pack mesh for \textit{k}-point sampling long  $\textless1\overline{1}0\textgreater$, $\textless11\overline{2}\textgreater$, and $\textless111\textgreater$ directions were used in the bulk supercell. In the case of PNHs, \textit{k}-point sampling along the direction normal to the interface plane is reduced to one. A planewave cutoff of 300 eV for the plane wave expansion of the wave functions was used.

\begin{table*}
\begin{center}
\caption{The bandgap of ZnSe and ZnTe in the bulk and in PNH geometries, calculated using HSE06. Percentage change in the bandgap under in-plane strain with respect to the equilibrium is indicated in parentheses. Experimental (measured) values of bandgap of ZnSe and ZnTe are also listed. Calculated lattice constant for ZnSe and ZnTe are, 5.74 \AA and 6.18 \AA, respectively.}
\begin{tabular} { p{1.3in} p{1.3in} p{1.0in} p{1.0in} p{1.0in} }
  \hline \hline
    \multirow{3}{*}{System} & \multirow{3}{*}{Strain \{planes\}} & \multicolumn{3}{c}{Bulk Bandgap (eV)} {PNH Bandgap (eV)} \\
 & & {Computed}  & {Measured} & {} \\
  \hline
\multirow{4}{*}{ZnSe} & Equilibrium  &   2.27 & 2.71  & 2.30\\
& +7.3\% $\{1\overline{1}0\}$  & 1.30(-43) & - & 1.30   \\
& +7.3\% $\{11\overline{2}\}$   & 1.44(-37)  & - & -   \\
& +7.3\% $\{111\}$ & 1.55(-32) & - & -    \\
\hline
 \multirow{4}{*} {ZnTe} & Equilibrium   & 2.03 & 2.39 & 2.10   \\
& -7.3\% $\{1\overline{1}0\}$ & 1.40(-31) &  - &  1.40  \\
& -7.3\% $\{11\overline{2}\}$   &2.12(+05) & - & -  \\
& -7.3\% $\{111\}$ & 1.73(-15) & - & -    \\
  \hline \hline
\end{tabular}
\label{table1}
\end{center}
\end{table*}

\section{Results and Discussions}

Table~\ref{table1} lists the bandgaps calculated using HSE06, of the equilibrium and in-plane strained bulk ZnSe and ZnTe.  Reported here is the maximum in-plane tensile and compressive strains in ZnSe and ZnTe, required to achieve coherency in the ZnSe/ZnTe PNHs. The equilibrium bandgap of ZnSe and ZnTe calculated using HSE06 are lower compared to the experimental value by 16.4\% and 15.0\%, respectively. The change in bandgap under the same in-plane strain values depends on the plane under consideration. Both ZnSe and ZnTe show the highest change in bandgap in the $\{1\overline{1}0\}$ plane under 7.3\% tensile and compressive strains, respectively. It should be noted that bandgap of ZnSe and ZnTe varies non-monotonically under compressive strain while it varies monotonically under tensile strain, as has been observed under strain in the [111] direction.\cite{4,10}

Now, we report on the effect of in-plane strains on the bandgap and band offset, in the PNH, with $\{1\overline{1}0\}$ interfacial plane orientation. The choice of interfacial plane was based on several reasons: 1) the interfacial plane orientation induces zero polarization,\cite{non-polar, graphatization} 2) free surfaces do not create new energy states, (which is crucial in explicitly determining the effect of epitaxial strains on the bandgap and the band offset in PNH) and 3) in-plane strain in the interfacial plane shows maximum tunability of bandgap.

\begin{figure}
\includegraphics[,width=3.0in]{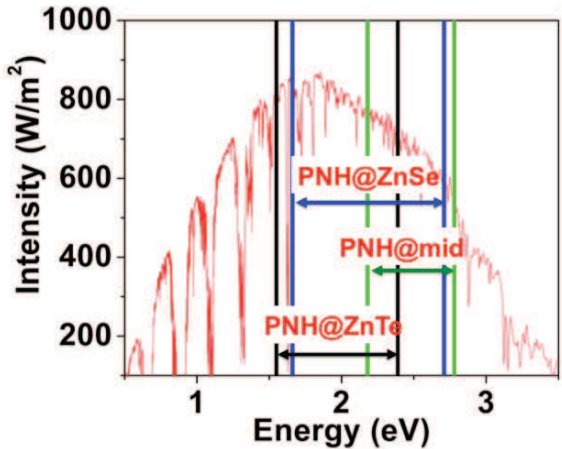}
\caption {(Color online) Experimental (estimated by scaling calculated values) bandgaps of various PNHs overlaid on solar spectra irradiance (ASTM G173-03, AM-1.5).\cite{solar-ape} For better comparison with the bandgap of PNHs, unit of wavelength of solar spectra is converted to eV.}
{\label{solar}}
\end{figure}

In Fig.~\ref{energy}, we plot the conduction band edge and valence band edge of the PNH as a function of distance from the interface. In the three PNH configurations considered, the in-plane lattice parameter was set equal to: a) the equilibrium lattice parameter of ZnSe, referred to as PNH@ZnSe (this induces 7.3\% in-plane compressive strain in ZnTe), b) average value of equilibrium lattice parameters of ZnSe and ZnTe, referred to as PNH@mid (this induces equal and opposite in-plane strain in ZnSe and ZnTe) and c) the equilibrium lattice parameter of ZnTe, referred to as PNH@ZnTe (this induces 7.3\% in-plane tensile strain in ZnSe). Band edges of PNHs were calculated using the layer decomposed density of states (LDOS) projected onto atomic planes.\cite{17,17-1} The LDOS for each layer is obtained by summing the density of states (DOS) projected from each atom in a layer and dividing by the total number of ZnSe or ZnTe pairs in that layer. In computing the LDOS, the energy eigenvalues are smeared with Gaussians of a width of 0.05 eV. 

For all PNHs configurations the bandgap varies non-monotonically across the interface due to bending of band edges. For PNH@ZnSe, the bandgap varies from 1.4 eV in ZnTe to 2.3 eV in ZnSe and for PNH@ZnTe, the bandgap varies from 2.0 eV in ZnTe to 1.3 eV in ZnSe. The presence of the range of bandgaps in nano-heterostructures such as ZnSe/ZnTe core/shell quantum dots is reflected by smooth absorption across a range of wavelength with no exciton absorption peak.\cite{b-3}  In all configurations the bandgap of ZnSe and ZnTe in PNH far from the interface have the same value as calculated in the bulk, thus not displaying artifacts related to quantum confinement effects. 

It is interesting to note that conduction band bending is always confined to ZnTe and valence band bending in ZnSe, revealing greater localization of the valence band edges to ZnTe and conduction band edges to ZnSe. A similar behavior was observed in Type-I, Si/SiO$_{2}$ heterostructure, where conduction band bending and valence band bending are confined in the same material system while in Type-II, CdSe/CdTe heterostructure, conduction band bending and valence band bending are confined in different parts of heterostructure.\cite{bending-cdse,bending-si}  

Strain not only affect the bandgap but also the absolute value of energy levels of both conduction and valence bands.\cite{slab-m} This leads to the valence band offset varying from 1.0 eV to 0 eV and the conduction band offset varying from 0.1 eV to 1.0 eV as the in-plane lattice parameters of the PNH is changed from the equilibrium lattice parameter of ZnSe to ZnTe. This is in agreement with previous experimental results. Strain tunable spectral range were observed in ZnTe/ZnSe core/shell quantum dots due to change in the bandgap and band edges as thickness of shell increases,\cite{b-2} a similar observation was reported in CdTe/ZnSe\cite{b-1} core shell quantum dots.

We estimate the range of bandgaps that should be observed experimentally in such heterostructures. The HSE06 calculated bandgaps (which are underestimated compared to experiment values) of ZnSe and ZnTe were increased by 16.4\% and 15.0\% to reproduce the experimental bandgap values. Fig.~\ref{solar} shows range of bandgaps in PNHs overlaid on the solar spectrum. Change in bandgap can be confirmed experimentally by measuring fluorescence from nanoheterostructures. Increase in the lower bound of the bandgap from PNH@ZnTe or PNH@ZnSe to PNH@mid will result in blue shift of peak intensity.

ZnSe/ZnTe heterostructures could be a suitable candidate for efficient solar energy capture because of: 1) large range of smoothly varying bandgap in the solar spectrum, and 2) efficient separation of electron hole pair due to Type-II band offset. A few nanometer layer of ZnSe expitaxially grown on thick ZnTe should be best candidate. This architecture shows a bandgap of 1.56 eV, which should result in Shockley-Queisser efficiency of 31\%.\cite{s-q}

\section{Conclusions}

In conclusion, the impact of epitaxial strains in ZnSe/ZnTe nano-heterostructures on the electronic properties have been studied using first-principles computations. We found that the bandgaps and the band offsets in heterostructure can be controlled by adequate strain in each component. This opens the possibility of tuning the bandgap and band offset of such heterostructures in the form of core/shell nanowire and thin films, by appropriately choosing the thickness of each component (which in turn controls strains). Strain in heterostructures may thus give a powerful degree of freedom for the rational design of optoelectronic devices.

\section*{Acknowledgements}
Financial support of this work through a grant from the National Science Foundation (NSF) and computational support through a NSF Teragrid Resource Allocation are acknowledged. Authors would like to acknowledge a critical reading of the manuscript by Ghanshyam Pilania and Arun Mannodi-Kanakkithodi.

\bibliographystyle{aipauth4-1}

\end{document}